%% file: anonymous-submission-latex-2026.tex
\documentclass[letterpaper]{article} 
\usepackage{aaai2026}  
\usepackage{times}  
\usepackage{helvet}  
\usepackage{courier}  
\usepackage[hyphens]{url}  
\usepackage{graphicx} 
\usepackage{natbib}  
\usepackage{caption} 
\usepackage{algorithm}
\usepackage{multirow}
\usepackage{amsmath}
\usepackage{amsfonts}
\usepackage{booktabs}
\usepackage{xcolor}
\usepackage{algpseudocode}
\usepackage{subcaption}
\usepackage{enumitem} 
\usepackage{graphicx}
\usepackage{booktabs} 
\usepackage{pdfpages} 
\usepackage{newfloat}
\usepackage{listings}
\DeclareCaptionStyle{ruled}{labelfont=normalfont,labelsep=colon,strut=off} 
\floatstyle{ruled}
\newfloat{listing}{tb}{lst}{}
\floatname{listing}{Listing}
\title{Beyond Negative Transfer: Disentangled Preference-Guided Diffusion for Cross-Domain Sequential Recommendation}

\author {
    Xiaoxin Ye\textsuperscript{\rm 1}, 
    Chengkai Huang\textsuperscript{\rm 1, \rm 2},
    Hongtao Huang\textsuperscript{\rm 1},
    Lina Yao\textsuperscript{\rm 1, \rm 3}
}
\affiliations {
    \textsuperscript{\rm 1}University of New South Wales,
    \textsuperscript{\rm 2}Macquarie University,
    \textsuperscript{\rm 3}CSIRO's Data61\\
    xiaoxin.ye@unsw.edu.au, chengkai.huang@mq.edu.au, hongtao.huang@unsw.edu.au,  lina.yao@data61.csiro.au
}

\begin{document}
\maketitle
\input{Content/01-Abstract}
\input{Content/02-Introduction}
\input{Fig/fig2_code}
\input{Content/03-Preliminary}

\input{Content/04-Method}

\input{Fig/PerformanceTable}
\input{Content/05-Experiment}
\input{Content/06-Relatedwork}
\input{Content/07-Conclusion}

%



\bibliography{aaai2026.bib}


\end{document}


\section{Technical Appendix}
Here is the detailed version of the DPGDiff inference process:
\input{Fig/Algo_Inference}

\subsection{Implementation Details}

\subsubsection{Dataset and Preprocessing}

We adopt the same datasets and preprocessing strategies as C2DREIF\cite{C2DREIF} , including the Amazon datasets. Two cross-domain scenarios are constructed: Food-Kitchen and Movie-Book. Following prior work, we exclude users with fewer than ten interactions and ensure each cross-domain sequence contains at least three items from each domain. Data is split into training, validation, and test sets using a leave-one-out protocol.

\subsubsection{Training Configuration}
\begin{itemize}
    \item Optimizer: Adam
    \item Learning Rate: 0.001
    \item Batch Size: 512
    \item Embedding Dimension: 256
    \item Heads:1
    \item Number of Layers: 2 for encoder and 1 for decoder
    \item Max Sequence Length: 15
    \item Epochs: 100
    \item Diffusion Steps: 50
    \item Hardware: NVIDIA L40S GPU (48 GB)
\end{itemize}

\subsection{Training Strategies}

To ensure stable convergence and effective representation learning, we adopt a multi-stage training schedule.

\begin{itemize}
    \item \textbf{Warm-up Phase:} We apply a linear learning rate warm-up over the first two training epochs. This stabilizes early training dynamics and mitigates gradient explosion, particularly for self-attention and cross-attention modules. During this phase, the model is trained using standard sequential recommendation objectives (cross-entropy loss), without contrastive or diffusion components.
    
    \item \textbf{Cosine Annealing with Full Loss:} After the warm-up, we switch to a cosine annealing schedule to gradually reduce the learning rate, promoting smoother convergence and better generalization. In this phase, we introduce the full loss configuration, including tri-view contrastive learning and diffusion-based denoising objectives. The contrastive loss aligns single-domain and cross-domain views, while the diffusion loss enhances temporal modeling and robustness.
\end{itemize}

\begin{table*}[htbp]
  \centering
  \small
  \setlength{\tabcolsep}{4pt}
  \begin{tabular}{l r r r r rr r r r r}
    \toprule
& \multicolumn{5}{c}{\textbf{Movie}} & \multicolumn{5}{c}{\textbf{Book}} \\
    \cmidrule(lr){2-6} \cmidrule(lr){7-11}
Algorithm  & MRR & N@5 & N@10 & H@5 & H@10 & MRR & N@5 & N@10 & H@5 & H@10 \\
    \midrule
    SAS & 3.79 & 3.23 & 3.69 & 3.99 & 5.20 & 1.81 & 1.41 & 1.71 & 1.83 & 2.75 \\
    SRGNN & 3.85 & 3.27 & 3.78 & 4.19 & 5.81 & 1.78 & 1.40 & 1.66 & 1.90 & 2.72 \\
    TCSR & 3.71 & 3.71 & 4.41 & 5.18 & 7.36 & 2.15 & 2.19 & 2.50 & 2.98 & 3.92 \\
    DiffuRec & 5.65 & 5.96 & 6.94 & 8.14 & 11.17 & 2.30 & 2.42 & 2.89 & 3.37 & 4.80 \\
    CGRec & 4.37 & 4.46 & 5.33 & 5.81 & 8.54 & 1.56 & 1.64 & 1.94 & 2.25 & 3.18 \\
    PSJNet & 4.63 & 4.06 & 4.76 & 5.30 & 7.53 & 2.44 & 2.07 & 2.35 & 2.58 & 3.28 \\
    MIFN & 5.05 & 4.41 & 5.20 & 5.51 & 8.29 & 2.51 & 2.12 & 2.31 & 2.46 & 3.07 \\
    C2DSR & 5.54 & 4.76 & 5.76 & 6.47 & 9.55 & 2.55 & 2.17 & 2.45 & 2.84 & 3.75 \\
    PCDSR & 5.77 & 4.93 & 5.79 & 6.21 & 8.89 & 2.44 & 2.03 & 2.31 & 2.60 & 2.31 \\
    EAGCL & 5.69 & 5.13 & 6.01 & 6.01 & 8.20 & 2.74 & 2.02 & 2.30 & 2.67 & 3.50 \\
    LEAGCN & 5.58 & 4.15 & 5.59 & 6.41 & 8.67 & 2.57 & 2.08 & 2.26 & 2.41 & 3.67 \\
    ABXI & 5.07 & 5.29 & 6.31 & 7.23 & 10.41 & 1.75 & 1.82 & 2.18 & 2.48 & 3.61 \\
    C2DREIF & 6.70 & 5.80 & 7.14 & 7.93 & 12.08 & 2.80 & 2.37 & 2.71 & 3.03 & 4.11 \\
    DREAM & \underline{8.74} & \underline{9.52} & \underline{10.58} & \underline{13.21} & \underline{16.49} & \underline{4.08} & \underline{4.24} & \underline{4.87} & \underline{5.58} & \underline{7.34} \\
    \midrule
    \textbf{Our} & \textbf{10.95*} & \textbf{11.60*} & \textbf{13.34*} & \textbf{15.68*} & \textbf{21.15*} & \textbf{4.92*} & \textbf{5.16*} & \textbf{5.79*} & \textbf{6.67*} & \textbf{8.63*} \\
    \midrule
    \textbf{Imp. (\%)} & +25.29 & +21.85 & +26.09 & +18.70 & +28.26 & +20.59 & +21.70 & +18.89 & +19.53 & +17.57 \\
    \bottomrule
  \end{tabular}
  \begin{minipage}{0.75\linewidth}
  
  \caption{Performance comparison on Movie and Book datasets. The best results are highlighted in \textbf{bold}, and the second-best are \underline{underlined}. \textbf{Imp.} indicates the relative improvement percentage of our model over the strongest baseline, * represents statistical significance confirmed via a t-test at $p < 0.05$ compared to the best-performing baseline.}
  \label{tab:perf_movie_book}
  \end{minipage}
  
\end{table*}

\begin{table*}[htbp]
  \centering
  \small
  \setlength{\tabcolsep}{4pt}
  \begin{tabular}{l r r r r rr r r r r}
    \toprule
 & \multicolumn{5}{c}{\textbf{Food}} & \multicolumn{5}{c}{\textbf{Kitchen}} \\
    \cmidrule(lr){2-6} \cmidrule(lr){7-11}
Algorithm & MRR & N@5 & N@10 & H@5 & H@10 & MRR & N@5 & N@10 & H@5 & H@10 \\
    \midrule
    SAS & 7.30 & 6.90 & 7.79 & 8.92 & 11.68 & 3.79 & 3.35 & 3.93 & 4.78 & 6.62 \\
    SRGNN & 7.84 & 7.58 & 8.35 & 9.88 & 12.27 & 4.01 & 3.47 & 4.13 & 4.80 & 6.84 \\
    TCSR & 7.25 & 7.17 & 8.26 & 9.24 & 12.62 & 4.24 & 3.22 & 5.66 & 4.79 & 5.66 \\
    DiffuRec & 7.85 & 8.46 & 9.13 & 11.10 & 13.21 & 3.60 & 3.90 & 4.55 & 5.62 & 7.67 \\
    CGRec & 8.19 & 8.73 & 9.43 & 11.21 & 13.40 & 3.99 & 4.25 & 4.86 & 5.81 & 7.70 \\
    PSJNet & 8.33 & 8.07 & 8.77 & 10.28 & 12.45 & 4.10 & 3.68 & 4.32 & 5.17 & 7.15 \\
    MIFN & 8.55 & 8.28 & 9.01 & 10.43 & 12.71 & 4.09 & 3.57 & 4.29 & 4.86 & 7.08 \\
    C2DSR & 8.91 & 8.65 & 9.71 & 11.24 & 14.54 & 4.65 & 4.16 & 4.94 & 5.74 & 8.18 \\
    PCDSR & 9.87 & 9.57 & 10.72 & 12.34 & 15.94 & 4.78 & 4.37 & 5.08 & 6.06 & 8.27 \\
    EAGCL & 9.93 & 9.05 & 10.22 & 12.54 & 15.07 & 4.89 & 4.28 & 5.16 & 6.22 & 8.37 \\
    LEAGCN & 9.14 & 8.76 & 9.85 & 12.70 & 15.85 & 4.78 & 4.12 & 5.28 & 6.07 & 8.58 \\
    ABXI & 9.19 & 9.89 & 10.50 & 12.74 & 14.63 & 4.70 & 5.10 & 5.75 & 7.12 & 9.12 \\
    C2DREIF & \underline{10.17} & 10.06 & 10.93 & 12.85 & 15.54 & 4.98 & 4.52 & 5.31 & 6.22 & 8.67 \\
    DREAM & 10.00 & \underline{10.75} & \underline{11.74} & \underline{14.27} & \underline{17.29} & \underline{5.04} & \underline{5.44} & \underline{6.37} & \underline{7.85} & \underline{10.73} \\
    \midrule
    \textbf{Our} & \textbf{11.20*} & \textbf{12.01*} & \textbf{13.07*} & \textbf{15.75*} & \textbf{19.04*} & \textbf{5.56*} & \textbf{6.02*} & \textbf{7.01*} & \textbf{8.65*} & \textbf{11.74*} \\
    \midrule
    \textbf{Imp. (\%)} & +10.13 & +11.72 & +11.33 & +10.37 & +10.12 & +10.32 & +10.66 & +10.05 & +10.19 & +9.41 \\
    \bottomrule
  \end{tabular}
  
  \begin{minipage}{0.73\linewidth}
\caption{Performance comparison on Food and Kitchen datasets. The best results are highlighted in \textbf{bold}, and the second-best are \underline{underlined}. \textbf{Imp.} indicates the relative improvement percentage of our model over the strongest baseline, * represents statistical significance confirmed via a t-test at $p < 0.05$ compared to the best-performing baseline.}
\label{tab:perf_food_kitchen}
\end{minipage}
\end{table*}




    
    


\subsection{Extended Baselines}

To provide a more comprehensive evaluation of our proposed DPG-Diff model, we conducted further comparisons with several additional baselines beyond those reported in the main paper. These models are categorized into sequential and cross-domain sequential recommendation.

\textbf{Sequential Recommendation}
\begin{itemize}
    \item \textbf{SAS} \cite{SASrec} models based on self-attention to capture user sequential behaviors.
    \item \textbf{SRGNN} \cite{SRGNN} leverage graph network to capture user sequential behaviors.
    \item \textbf{TCSR} \cite{TiSASRec} An extension algorithm of CoSeRec\cite{CoSeRec} using SASRec as a foundational model with a time-aware self-attention mechanism and contrastive learning.
    \item \textbf{DiffuRec}\cite{diff4serec} A strong baseline using diffusion model for sequential recommendation
\end{itemize}

\textbf{Cross-domain Sequential Recommendation}
\begin{itemize}
    \item \textbf{$\pi$-Net}\cite{pinet} and \textbf{PSJNet}\cite{PSJNet}: Integrate cross-domain recommendation and sequential recommendation techniques, using fine-grained gate mechanisms for information transfer and individual cross-domain knowledge transfer.
    \item \textbf{MIFN}\cite{MIFN}: Constructs interaction graphs between different domain items to facilitate cross-domain information transfer.
    \item \textbf{C$^{2}$DSR}\cite{C2DSR}: Addresses single-domain and cross-domain issues by merging infomax objectives to capture cross-domain relevance.
    \item \textbf{PCDSR}~\cite{PCDSR}: Incorporates proxy items and temporal intervals. 
    \item \textbf{EAGCL}~\cite{EAGCL}: Uses external attention and memory-sharing MLPs. 
    \item \textbf{LEAGCN}~\cite{LEAGCN}: Captures collaborative filtering signals across domains. 
    \item \textbf{ABXI}~\cite{ABXI}: Introduces invariant interest adaptation using LoRA-based modules.
    \item \textbf{C2DREIF}~\cite{C2DREIF}: Combines contrastive learning with disentangled representations. 
    \item \textbf{DREAM} ~\cite{DREAM}: A recent CDSR model leveraging decoupled extraction and contrastive learning mechanisms.
\end{itemize}

\subsection{Performance compare}
Table (\ref{tab:perf_movie_book} and \ref{tab:perf_food_kitchen}) are the full performance comparison table including some algorithms which were excluded from the main table due to space constraints. The extended results, presented in Table (\ref{tab:perf_movie_book} and \ref{tab:perf_food_kitchen}), consistently reinforce the effectiveness of DPG-Diff across all metrics and domains. Notably, DPG-Diff achieves superior performance even when compared against these stronger baselines, often outperforming them in cross-domain settings. This expanded evaluation further validates the robustness and generalization capability of our model.

\subsection{Reproducibility}
To facilitate reproducibility, we have included a ZIP archive containing the full source code and one model object for the Food-Kitchen experiments. The shared model is stored in float16 precision to reduce the memory footprint due to the file upload limit. Despite the reduced precision, our evaluations show minimal performance degradation compared to the original float32 version, demonstrating the robustness of DPG-Diff under quantization. The provided scripts support direct inference and evaluation, and are structured to mirror the experimental pipeline described in the main paper.

\appendix
\bibliography{AnonymousSubmission/LaTeX/aaai2026}


%% file: Content/01-Abstract.tex
\begin{abstract}

Cross-Domain Sequential Recommendation (CDSR) leverages user behaviors across domains to enhance recommendation quality. However, naive aggregation of sequential signals can introduce conflicting domain-specific preferences, leading to negative transfer. While Sequential Recommendation (SR) already suffers from noisy behaviors such as misclicks and impulsive actions, CDSR further amplifies this issue due to domain heterogeneity arising from diverse item types and user intents. The core challenge is disentangling three intertwined signals: domain-invariant preferences, domain-specific preferences, and noise. Diffusion Models (DMs) offer a generative denoising framework well-suited for disentangling complex user preferences and enhancing robustness to noise. Their iterative refinement process enables gradual denoising, making them effective at capturing subtle preference signals. However, existing applications in recommendation face notable limitations: sequential DMs often conflate shared and domain-specific preferences, while cross-domain collaborative filtering DMs neglect temporal dynamics, limiting their ability to model evolving user preferences. To bridge these gaps, we propose \textbf{DPG-Diff}, a novel Disentangled Preference-Guided Diffusion Model, the first diffusion-based approach tailored for CDSR, to or best knowledge. DPG-Diff decomposes user preferences into domain-invariant and domain-specific components, which jointly guide the reverse diffusion process. This disentangled guidance enables robust cross-domain knowledge transfer, mitigates negative transfer, and filters sequential noise. Extensive experiments on real-world datasets demonstrate that DPG-Diff consistently outperforms state-of-the-art baselines across multiple metrics.

\end{abstract}

%% file: Content/02-Introduction.tex
\begin{figure}[h]
    \centering
        \centering
        \includegraphics[width=0.47\textwidth]{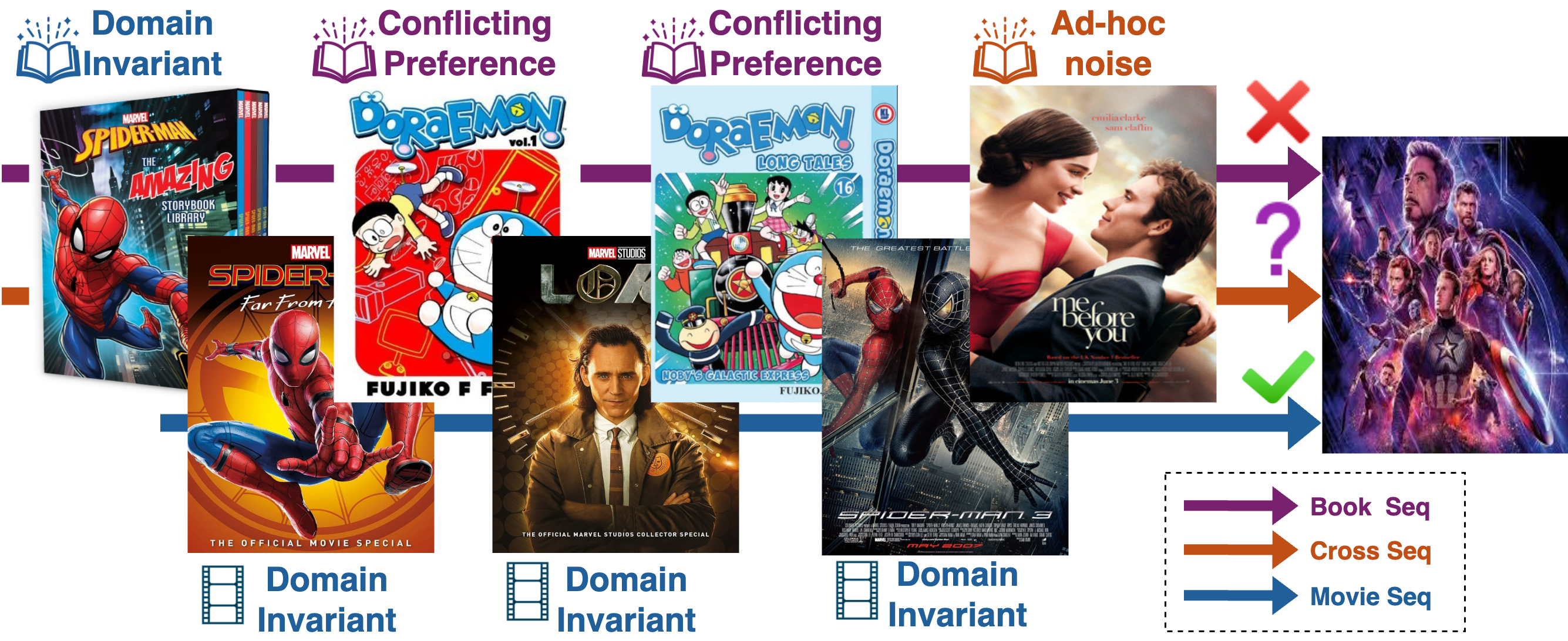}
        \caption{The complexity of user behavior sequences in cross-domain scenario, including three types of user preferences: domain-invariant, conflicting, and ad-hoc noise. These mixed signals create challenges for models to learn user intent across domains.
        }
        \label{fig:sub1}
    \label{fig:combined}
\end{figure}

\section{Introduction}
In an era of fragmented digital ecosystems, users interact with diverse content across multiple platforms, creating a pressing need for recommender systems that can understand and adapt to cross-domain, sequential user behavior \cite{huang2024foundation, huang2025survey, huang2023dual, huang2023modeling, zhou2023contrastive}. 

While Cross-Domain Recommendation (CDR) models static user interactions across domains, Cross-Domain Sequential Recommendation (CDSR) extends this by capturing temporal user behavior. This introduces new challenges: preferences are dynamic, domain-dependent, and often noisy, sparse, or inconsistent.

Consider the user sequence in Fig.~\ref{fig:sub1}: the user begins with a Spider-Man book, followed by multiple Spider-Man movies. This suggests a shared interest in Marvel content.  However, this signal is sparse in the book domain and disrupted with unrelated purchases like Doraemon and other books. This example highlights three key challenges:

\textbf{Challenge 1: Capturing domain-invariant sequential preference.} Users often exhibit long-term interests (e.g., Marvel) that span domains. These signals are diluted by domain-specific signals (e.g., Doraemon books), forming sequences like Marvel $\rightarrow$ Doraemon $\rightarrow$ Marvel. Traditional sequential models, which prioritize recent interactions~\cite{sr_survery}, struggle with such long-range, cross-domain dependencies. Prior works (e.g., C2DSR~\cite{C2DSR}) attempt to learn shared representations using global structures but often assume clean alignment, which fails under real-world heterogeneity and inconsistency.

\textbf{Challenge 2: Negative transfer from contextually conflicting domain-specific behaviors.}  Domain-specific interactions can conflict with domain-invariant preference, further complicating the CDSR task. In the same sequence, the user only purchase Doraemon books (a likely purchase for a child), but shows no interest in Doraemon elsewhere. This reflects a domain-specific intent that does not generalize across domains. When such behaviors are interleaved with domain-invariant signals, it introduces conflicting patterns. Naively merging such sequences can confuse the model and degrade performance\cite{negative_transfer}. Models like CGRec~\cite{negative_transfer} attempt to mitigate this via game theory and contrative learning, but balancing generalization and specificity remains a core challenge\cite{transferability}.

\textbf{Challenge 3: Contextually irrelevant or impulsive actions} In CDSR, users often exhibit contextually irrelevant or impulsive actions in one domain that obscure their core preferences in another. For example, as shown in Fig.~\ref{fig:sub1}, the user purchases a romantic book (‘Me Before You') as a one-off gift, which introduces noise into the sequence. Unlike standard SR, where behaviors are confined to a single domain, CDSR must disentangle domain-invariant preference from domain-specific and ad-hoc signals simultaneously. This is especially challenging due to behavioral inconsistency\cite{negative_transfer}. Without robust disentanglement, models risk capturing spurious correlations, increasing the risk of negative transfer, and degrading the model quality.

These challenges underscore the need for a model capable of \textbf{robustly disentangling and selectively retaining meaningful signals across domains}, while also capturing the \textbf{temporal dynamics} of user behavior. Traditional sequential models and existing CDSR approaches often fall short due to their reliance on \textit{recent interactions}, \textit{rigid fusion strategies}, or \textit{domain-specific assumptions}, which limit their ability to generalize across heterogeneous domains.

Diffusion Models (DMs) offer a compelling solution. Originally developed for generative tasks in vision and language~\cite{YangZSHXZZCY24}, DMs operate through a two-step process: a forward diffusion that gradually corrupts data with noise, and a reverse denoising process that reconstructs the original signal. This iterative framework is inherently well-suited for disentangling complex, noisy, and multi-source inputs, making it a natural fit for the challenges. DiffuRec~\cite{diff4serec} has demonstrated that DMs can effectively capture sequential dependencies in single-domain settings by modeling user behavior as a generative denoising process. Nevertheless, DMCDR~\cite{DMCDR} and CDCDR~\cite{CDCDR} introduced preference-guided diffusion for cross-domain recommendation. However, these models are designed for general Cross-Domain Recommendation (CDR) and operate under the assumption of static user interactions. As a result, they fail to incorporate the temporal dynamics and user sequential behavior, which are central to the CDSR setting.

Extending diffusion to \textbf{Cross-Domain Sequential Recommendation (CDSR)} requires more than denoising. It demands a mechanism that can \textbf{semantically align and temporally guide} the generation process. To simultaneously address those issue, we propose Disentangled Preference-Guided Diffusion (\textbf{DPG-Diff}), the first diffusion-based model tailored for CDSR, to our best knowledge. DPG-Diff introduces a generative view of user behavior, modeling preferences through a guided denoising process that naturally handles behavioral noise and domain heterogeneity. 

At its core, DPG-Diff employs a disentanglement module to separate user preferences into domain-specific components and domain-specific components. These disentangled signals jointly guide the reverse diffusion process, enabling the model to selectively reconstruct meaningful user representations while suppressing contextually irrelevant or conflicting signals. This structured guidance not only mitigates negative transfer but also enhances the model’s ability to capture long-range, cross-domain sequential patterns.



\paragraph{Our main contributions are summarized as follows:}
\begin{itemize}

    \item To our best knowledge, we propose the first diffusion-based model tailored for CDSR, Disentangled Preference-Guided Diffusion model(\textbf{DPG-Diff}), introducing a generative paradigm that captures complex and dynamic user preferences across domains. 
    \item We design a novel disentangled sequential preference-guided diffusion mechanism that separates domain-invariant and domain-specific user preferences. 
    \item We embed disentangled preferences into the reverse diffusion process, enabling robust cross-domain transfer while suppressing noisy or conflicting behaviors.
    \item We conduct extensive experiments on multiple real-world datasets, demonstrating that DPG-Diff consistently outperforms SOTA CDSR models.

\end{itemize}

%% file: Fig/fig2_code.tex
\definecolor{noise}{HTML}{E2725B}
\definecolor{interest}{HTML}{3B82F6}
\definecolor{domain_interest}{HTML}{EC4899}
\begin{figure*}[htbp]
\center
    \centering
    \includegraphics[width=1\textwidth, trim=55bp 50bp 55bp 135bp, clip]{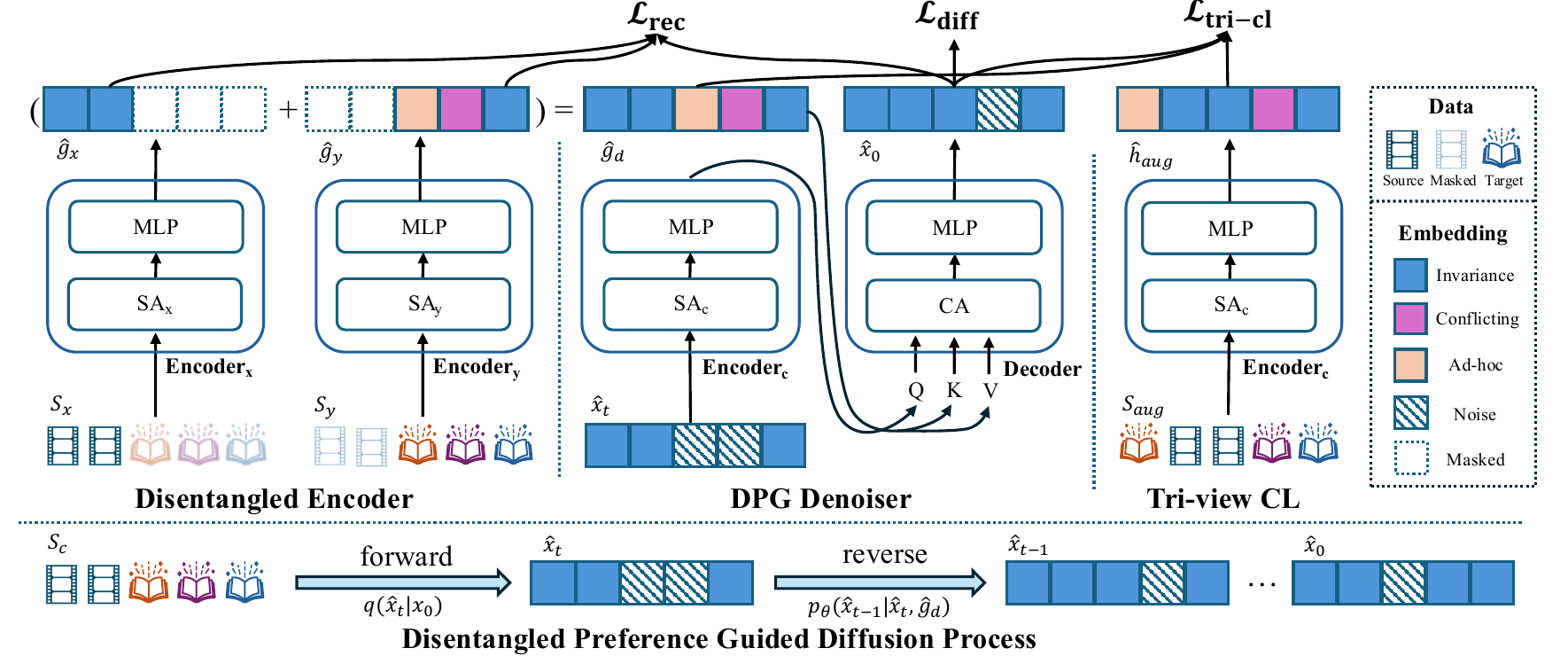}
\caption{\textbf{DPG-Diff} integrates a \textbf{Disentangled Encoder}, a Disentangled Preference Guided Denoiser (\textbf{DPG Denoiser}) and \textbf{Tri-view CL} within a diffusion recommendation architecture. The input is a cross-domain sequence \( S_c \), composed of $S_x$ for domain x, and $S_y$ for domain y. $S_c$ may contains three disentangled preference: domain-invariant, conflicting, and ad-hoc noise. The Disentangled Encoder extracts representations from each domain, which guide the reverse diffusion process. The tri-view CL aims to enhance alignment between cross-domain and domain-specific view. \textbf{Diffusion Process} performs forward corruption and guided reverse denoising to produce $\hat{x}_0$, an estimated user interests.}
    \label{fig:2}
\end{figure*}


%% file: Content/03-Preliminary.tex
\section{Preliminary }
This section provides background on vanilla diffusion models\cite{HoJA20}, and problem formulation.
\subsubsection{Vanilla Diffusion Models} (DMs) typically involve forward and reverse processes, formulated as two Markov chains to model the underlying data generating distribution.

\noindent\textbullet{} \textbf{Forward process}.  Given a sample $x_0 \sim q(x_0)$, DMs progressively add Gaussian noise over $T$ steps via a Markov chain $\{x_0, x_1, \dots, x_T\}$. Each transition is defined as:
\small
\begin{align}
    q(x_t \mid x_{t-1}) &= \mathcal{N}(x_t; \sqrt{1 - \beta_t} \, x_{t-1}, \, \beta_t \mathbf{I}),
\end{align}
\normalsize
where $\beta_t \in (0, 1)$ controls the noise variance at step $t$. The full forward process is defined as:
\small
\begin{align}
q(x_{1:T} \mid x_0) &= \prod_{t=1}^{T} q(x_t \mid x_{t-1}),
\end{align}
\normalsize
 With the reparameterization trick\cite{reparameter}, $x_t$ can be sampled directly from $x_0$:
\small
\begin{align}
\label{forward}
        q(x_t \mid x_0) &= \mathcal{N}(x_t; \sqrt{\bar{\alpha}_t} \, x_0, (1 - \bar{\alpha}_t) \mathbf{I}), where\\
        x_t &= \sqrt{\bar{\alpha}_t} \, x_0 + \sqrt{1 - \bar{\alpha}_t} \, \epsilon, \quad \epsilon \sim \mathcal{N}(0, \mathbf{I}),
\end{align}
\normalsize

with $\alpha_t = 1 - \beta_t$ and $\bar{\alpha}_t = \prod_{s=1}^{t} \alpha_s$. As $T \to \infty$, the distribution of $x_T$ converges to standard Gaussian noise. 

\noindent\textbullet{} \textbf{Reverse process} focuses on recovering the sample $x_0$ through a progressive denoising process modeled as a Markov chain. Each denoising transition $x_t\rightarrow x_{t-1}$ is parameterized by a linear Gaussian transition. The arithmetic mean $\mu_\theta$ and standard deviation $\Sigma_\theta$ are calculated by a denoising DNN $f_\theta(e_t, t)$. Therefore, the transition with trainable parameters $\theta$, formulated as: 
\small
\begin{align} \label{equ:reverse_transition}
    p_\theta(x_{t-1}|x_t)=\mathcal{N}(x_{t-1};\mu_\theta(x_t, t),\Sigma_\theta(x_t, t)),
\end{align}
\normalsize

\noindent\textbullet{} \textbf{Optimization} The goal of training diffusion models is to approximate the true data distribution \( p_{\theta}(x) \) by minimizing the negative log-likelihood. Following prior works \cite{dhariwal2021diffusion, DMCDR}, this objective is typically simplified as a mean squared error (MSE) loss:

\small
\begin{equation}\label{equ: loss_elbo_z_continous_final}
\mathcal{L}_{\text{elbo}}(x_0) = \mathbb{E}_{q(x_{1:T}|x_0)} [\sum_{t=1}^T \left\| f_\theta(x_t, t) - x_0 \right\|^2 ],
\end{equation}
\normalsize

This formulation encourages the model \( f_\theta \) to learn how to denoise the corrupted representations \( x_t \) at each time step, ultimately reconstructing the original input \( x_0 \).

\noindent\textbullet{} \textbf{Inference} Inference proceeds by sampling a Gaussian noise \( x_T \sim \mathcal{N}(0, I) \) and applying the learned reverse process \( p_{\theta}(x_{t-1}|x_t) \) iteratively to generate the final output \( \hat{x} \).

\subsection{Problem Formulation} 
In this paper, we focus on \textbf{Cross-Domain Sequential Recommendation (CDSR)}, where each user interacts with two domains, denoted as domain $\mathcal{X}$ and domain $\mathcal{Y}$. 

For a given user $u$, we define their cross-domain sequence as $s_c = [v_1, v_2, \dots, v_{|s_c|}]$, where each interaction $v_i \in \mathcal{X} \cup \mathcal{Y}$ and the sequence is ordered by interaction timestamps. From $s_c$, we derive domain-specific subsequences $s_x \subset s_c$ and $s_y \subset s_c$, containing only items from domains $\mathcal{X}$ and $\mathcal{Y}$.

\textbf{Objective.} Given the cross-domain sequence $s_c$ up to time $t$, the goal is to predict the next item $v_{t+1} \in \mathcal{X} \cup \mathcal{Y}$ that the user is most likely to interact with, formulated as: 
\small
\begin{align}
    v_{t+1} = \arg\max_{v \in \mathcal{X} \cup \mathcal{Y}} P(v \mid s_c),
\end{align}
\normalsize

%% file: Content/04-Method.tex
\section{Methodology}

This section introduces the proposed \textbf{DPG-Diff} framework for Cross-Domain Sequential Recommendation (CDSR). As shown in Fig.~\ref{fig:2}, DPG-Diff establishes a novel generative paradigm that integrates disentangled user preferences into a diffusion-based denoising process. The framework comprises three key components: a \textit{Disentangled Encoder}, a \textit{Preference-Guided Denoiser}, and \textit{Tri-view Contrastive Learning (Tri-view CL)}. Unlike prior approaches that statically or rigidly merge cross-domain signals, DPG-Diff constructs structured, domain-aware representations to guide a reverse diffusion process that reconstructs user behavior with high fidelity. Tri-view CL further refines these representations by leveraging multi-view alignment.

\subsection{Disentangled Encoder}

This module generates domain-specific and invariant representations by disentangling user preferences across domains, including an embedding and an encoder layer.

\subsubsection{Disentangled Embedding Layer}
We begin by constructing the full cross-domain sequence $s_c$, along with domain-specific sequences $s_x$ and $s_y$, preserving temporal order. An augmented sequence $s_{\text{aug}}$ is also generated for Tri-view CL. 

Each domain has its own item embedding $E_x, E_y \in \mathbb{R}^{d}$, and a shared positional embedding $\text{Pos} \in \mathbb{R}^{d}$ encodes temporal structure. For a sequence $ s_y = [y_1, y_2, \dots, y_t]$, the input representation is computed as $ h_y= \text{Emb}(s_y) + \text{Pos}(s_y)$.

\subsubsection{Sequential Encoder} To capture sequential dependencies and disentangle user preferences, we employ Transformer-based encoders for each domain independently, denoted as \( \text{Encoder}_x \) and \( \text{Encoder}_y \). Each encoder processes the input sequence representations \( h_x \) and \( h_y \), producing domain-specific outputs \( g_x \) and \( g_y \):
\small
    \begin{align}
        \label{encoder}
        g_x = \text{Encoder}_x(h_x), \quad g_y = \text{Encoder}_y(h_y),
    \end{align}
\normalsize

Each encoder consists of a self-attention layer (SA) to model long-range dependencies and capture contextual relationships within a sequence, followed by a MLP to project the attended features into disentangled latent spaces. This design enables the encoder to effectively separate domain-specific and domain-invariant signals.

As illustrated in Fig.~\ref{fig:2}, the resulting representations \( g_x \) and \( g_y \) encapsulate domain-specific semantics. We fuse them into a unified cross-domain representation \( g_d \), which serves as structured guidance for the reverse diffusion process. This fusion facilitates effective cross-domain knowledge transfer while mitigating negative transfer caused by conflicting or noisy signals.

\subsection{Disentangled Preference Guided Diffusion Process}

The reverse process in \textbf{DPG-Diff} is designed to iteratively refine corrupted user representations \( x_t \) into semantically rich embeddings \( \hat{x}_0 \) for downstream recommendation tasks. 


As illustrated in Fig.~\ref{fig:2}, the proposed Disentangled Preference-Guided Denoiser \textbf{(DPG Denoiser)} consists of two Transformer-based modules: a domain-invariant encoder \( \text{\textbf{Encoder}}_\textbf{c} \) and a cross-domain \textbf{Decoder}. The encoder captures high-level semantics from noisy input via self-attention (SA) and an MLP, modeling long-range dependencies and abstract patterns. The decoder employs cross-attention (CA) and an MLP to selectively fuse the encoded representation with structured guidance \( g_d \), synthesized from domain-specific signals \( g_x \) and \( g_y \):
\small
\begin{align}
\label{denoising}
    \hat{x}_0 = f_\theta(x_t, g_d) &= \text{Decoder}(\text{Encoder}_c(x_t), g_d),
\end{align}
\normalsize
The cross-attention mechanism is critical to align noisy inputs with semantically coherent signals, enabling the decoder to focus on relevant aspects of user preferences while filtering out noise. This selective fusion preserves domain-specific nuances and enhances cross-domain generalization. %

Unlike prior works that statically merge cross-domain signals, DPG-Diff dynamically adapts to the semantic structure of user behavior. Guided by both domain-specific and domain-invariant signals, the model reconstructs behaviorally consistent embeddings, promoting effective knowledge transfer while mitigating negative transfer and random noise. This design significantly improves robustness and recommendation accuracy in cross-domain settings.


By iteratively applying this denoising framework, DPG-Diff reconstructs user embeddings with high fidelity, guided by both domain-specific and domain-invariant signals. The fused guidance \( g_d \) acts as a semantic prior, enabling the model to effectively distinguish signal from noise and generate behaviorally consistent representations. This design not only promotes effective knowledge transfer across domains but also mitigates the risk of negative transfer and random noise, significantly enhancing the model’s robustness and generalization in cross-domain settings.

\subsection{DPG-Diff Training Objective}

Upon completing the diffusion process, the model produces four distinct user representations: (1) $\hat{x}_0$, the denoised cross-domain preference from the DPG Denoiser; (2) $\hat{g}_x$, the single-domain preference from Encoder$_x$; (3) $\hat{g}_y$, the single-domain preference from Encoder$_y$; and (4) $\hat{g}_d$, the fused single-domain preference derived from $\hat{g}_x$ and $\hat{g}_y$.

This multi-view representation framework enables DPG-Diff to capture nuanced user behaviors across domains. By disentangling and selectively fusing domain-specific signals, the model facilitates effective knowledge transfer while mitigating risks of negative transfer and noise interference. The structured guidance mechanism preserves semantic integrity and enhances adaptability to evolving user preferences. Consequently, DPG-Diff operates robustly in both single-domain and cross-domain settings, offering a unified and noise-resilient solution for sequential recommendation.

\textbf{Disentangled Recommendation Loss ($\mathcal{L}_{\text{rec}}$)} To ensure the Disentangled Encoder remains resilient to cross-domain noise, we decouple learning signals from single-domain and cross-domain perspectives. The overall recommendation loss is defined as:
\small
    \begin{align}
    \label{loss:rec}
    \mathcal{L}_{\text{rec}} &= \textstyle{\sum}_{d \in \{x, y\}} ( \mathcal{L}_d^c + \mathcal{L}_d^s ), where\\
    \mathcal{L}_d^c &= \text{CE}(\text{softmax}(\hat{x}_0^\top E_d), v_d),\\
    \mathcal{L}_d^s &= \text{CE}(\text{softmax}( \hat{g}_d^\top E_d ), v_d),
    \end{align}
\normalsize

Here, $\mathcal{L}_d^c$ and $\mathcal{L}_d^s$ denotes the cross-domain and single-domain recommendation loss, and CE donates the Cross-entropy loss.
\( E_d \) is the item embedding matrix for domain \( d \), and \( v_d \) is the ground-truth item.


\textbf{Tri-View Contrastive Loss ($\mathcal{L}_{\text{tri-cl}}$)}. To further align cross-domain and domain-specific representations, we introduce a tri-view contrastive learning objective. Each user is represented by three embeddings: (1)  $h_c = \hat{x}_0$, the denoised cross-domain embedding; (2) $h_d = \hat{g}_d$, the fused domain-aware embedding; and (3) \( h_{\text{aug}} = \text{Encoder}_c(s_{\text{aug}}) \), the augmented embedding derived from five sequence augmentations: Crop, Mask, Reorder, Substitute, and Insert. Intra-user pairs, $(h_c^i, h_d^i)$, $(h_c^i, h_{\text{aug}}^i)$, and $(h_d^i, h_{\text{aug}}^i)$, are treated as positives, while all other combinations are considered negatives. The contrastive loss is formulated via the objective:
\small
    \begin{align}
    \label{loss:cl}
    \mathcal{L}_{\text{tri-cl}} = -\log ( \frac{\exp(h_i^\top h_j)}{\sum_{h^{-} \in \mathcal{H}_i^-} \exp(h_i^\top h^{-})+ \exp(h_i^\top h_j) } ),
    \end{align}
\normalsize

This objective strengthens the coherence among different views of the same user while enforcing discriminative separation across users.

\input{Fig/Algo_Train}

\subsection{Optimization and Inference}
\label{sec:Optimization}
\textbf{Optimization} The overall training process is illustrated in Algorithm 1, using a mini-batch training approach. The training objective integrates three components:
\small
\begin{align}
\mathcal{L}_{\text{total}} = \mathcal{L}_{\text{diff}} + \mathcal{L}_{\text{rec}} + \mathcal{L}_{\text{tri-cl}},
\end{align}
\normalsize
where $\mathcal{L}_{\text{diff}} $ is the diffusion reconstruction loss, learning to reconstruct the clean item embedding \( x_0 \) from its noisy version \( x_t \) using mean squared error, using Eq. (\ref{equ: loss_elbo_z_continous_final}); $\mathcal{L}_{\text{rec}} $ is the disentangled recommendation loss, using Eq. (\ref{loss:rec}); $\mathcal{L}_{\text{tri-cl}} $ is the Tri-View Contrastive Loss using Eq.(\ref{loss:cl}).

\label{sec:Inference}
\textbf{Inference} During inference, given a user's interaction sequence, Diff-Rec first encodes the sequence using the Disentangled Encoder to obtain single-domain preferences \( \hat{g}_x \) and \( \hat{g}_y \), which are fused into a unified guidance representation \( \hat{g}_d \). This guidance is used by the DPG-Denoiser to iteratively reconstruct the cross-domain preference \( \hat{x}_0 \) from a randomly sampled noisy vector \( x_{t}\) at each step via Eq. (\ref{denoising})

Once the final cross-domain denoised representation $\hat{x}_0$ is obtained, the model generates prediction scores by integrating cross-domain (\( \hat{x}_0 \)) and domain-specific signals (\( \hat{g}_x, \hat{g}_y \)). For domain y, the scores are computed as:
\small
\begin{equation}
\label{eq:score}
\hat{p}_y = \text{softmax}(\hat{x}_0 + \hat{g}_y)^\top E_y ).
\end{equation}
\normalsize

$E_y$ denotes the item embedding matrix for domain $y$. The top-\(k\) items with the highest scores in $\hat{p}_y$ are selected as recommendations (analogously for domain x).

This inference mechanism allows DPG-Diff to seamlessly fuse disentangled preference guidance with diffusion-based reconstruction, enhancing recommendation accuracy and robustness across domains. More details are available in the supplementary material.


%% file: Fig/Algo_Train.tex
\begin{algorithm}[t]
\small
\label{training}
\caption{Training Procedure of DPG-Diff}
\begin{algorithmic}[1]
\Require Cross-domain $s_c$, item embeddings $E_x$, $E_y$
\Ensure Trained model parameters $\theta$
\For{each training batch}
    \State Extract domain-specific sequences $s_x$, $s_y$ from $s_c$
    \State Encode $s_x$, $s_y$ into $g_x$, $g_y$ via Eq.~(\ref{encoder})
    \State Fuse $g_x$, $g_y$ into disentangled guidance $g_d$
    \State Sample timestep $t \sim \mathcal{U}(1, T)$ and noise $\epsilon \sim \mathcal{N}(0, I)$
    \State Perform $q(x_t|x_0)$ in  Eq. (\ref{forward})
    \State Predict denoised embedding: $\hat{x}$ via Eq. (\ref{denoising})
    \State Compute diffusion loss: $\mathcal{L}_{\text{diff}}$ via Eq. (\ref{equ: loss_elbo_z_continous_final})
    \State Compute diffusion loss: $\mathcal{L}_{\text{rec}}$ via Eq. (\ref{loss:rec})
    \State Compute contrastive loss $\mathcal{L}_{\text{tri-cl}}$ via Eq. (\ref{loss:cl})
    \State Update parameters: $\theta \leftarrow \theta - \nabla_\theta (\mathcal{L}_{\text{diff}} +\mathcal{L}_{\text{rec}} + \mathcal{L}_{\text{tri-cl}})$ 
\EndFor
\end{algorithmic}
\end{algorithm}

%% file: Fig/PerformanceTable.tex
\begin{table*}[htbp]

  \setlength{\tabcolsep}{1mm}
  \small
\begin{tabular}{cl |ccc|cccccccccc|c}
    \toprule
    \textbf{Data} & \textbf{Metric} & \textbf{SAS} & \textbf{TCSR} & \textbf{Diffu} & \textbf{CGRec} & \textbf{PSJ} & \textbf{C$^2$DSR} & \textbf{PCDSR} & \textbf{EAGCL} & \textbf{LGCN} & \textbf{ABXI} & \textbf{C$^2$DREIF} & \textbf{DREAM} & \textbf{Our} & \textbf{Imp.} \\
    \midrule
    \multirow{5}{*}{\rotatebox{90}{\textbf{Movie}}} & MRR & 3.79 & 3.71 & 5.65 & 4.37 & 4.63 & 5.54 & 5.77 & 5.69 & 5.58 & 5.07 & 6.70 & \underline{8.74} & \textbf{10.95} & 25.3\% \\
     & N@5 & 3.23 & 3.71 & 5.96 & 4.46 & 4.06 & 4.76 & 4.93 & 5.13 & 4.15 & 5.29 & 5.80 & \underline{9.52} & \textbf{11.60} & 21.8\% \\
     & N@10 & 3.69 & 4.41 & 6.94 & 5.33 & 4.76 & 5.76 & 5.79 & 6.01 & 5.59 & 6.31 & 7.14 & \underline{10.58} & \textbf{13.34} & 26.1\% \\
     & H@5 & 3.99 & 5.18 & 8.14 & 5.81 & 5.30 & 6.47 & 6.21 & 6.01 & 6.41 & 7.23 & 7.93 & \underline{13.21} & \textbf{15.68} & 18.7\% \\
     & H@10 & 5.20 & 7.36 & 11.17 & 8.54 & 7.53 & 9.55 & 8.89 & 8.20 & 8.67 & 10.41 & 12.08 & \underline{16.49} & \textbf{21.15} & 28.2\% \\
\cmidrule(lr){2-16}
    \multirow{5}{*}{\rotatebox{90}{\textbf{Book}}} & MRR & 1.81 & 2.15 & 2.30 & 1.56 & 2.44 & 2.55 & 2.44 & 2.74 & 2.57 & 1.75 & 2.80 & \underline{4.08} & \textbf{4.92} & 20.5\% \\
     & N@5 & 1.41 & 2.19 & 2.42 & 1.64 & 2.07 & 2.17 & 2.03 & 2.02 & 2.08 & 1.82 & 2.37 & \underline{4.24} & \textbf{5.16} & 21.6\% \\
     & N@10 & 1.71 & 2.50 & 2.89 & 1.94 & 2.35 & 2.45 & 2.31 & 2.30 & 2.26 & 2.18 & 2.71 & \underline{4.87} & \textbf{5.79} & 18.9\% \\
     & H@5 & 1.83 & 2.98 & 3.37 & 2.25 & 2.58 & 2.84 & 2.60 & 2.67 & 2.41 & 2.48 & 3.03 & \underline{5.58} & \textbf{6.67} & 19.6\% \\
     & H@10 & 2.75 & 3.92 & 4.80 & 3.18 & 3.28 & 3.75 & 2.31 & 3.50 & 3.67 & 3.61 & 4.11 & \underline{7.34} & \textbf{8.63} & 17.5\% \\
    \midrule
    \multirow{5}{*}{\rotatebox{90}{\textbf{Food}}} & MRR & 7.30 & 7.25 & 7.85 & 8.19 & 8.33 & 8.91 & 9.87 & 9.93 & 9.14 & 9.19 & \underline{10.17} & 10.00 & \textbf{11.20} & 10.1\% \\
     & N@5 & 6.90 & 7.17 & 8.46 & 8.73 & 8.07 & 8.65 & 9.57 & 9.05 & 8.76 & 9.89 & 10.06 & \underline{10.75} & \textbf{12.01} & 11.7\% \\
     & N@10 & 7.79 & 8.26 & 9.13 & 9.43 & 8.77 & 9.71 & 10.72 & 10.22 & 9.85 & 10.50 & 10.93 & \underline{11.74} & \textbf{13.07} & 11.3\% \\
     & H@5 & 8.92 & 9.24 & 11.10 & 11.21 & 10.28 & 11.24 & 12.34 & 12.54 & 12.70 & 12.74 & 12.85 & \underline{14.27} & \textbf{15.75} & 10.4\% \\
     & H@10 & 11.68 & 12.62 & 13.21 & 13.40 & 12.45 & 14.54 & 15.94 & 15.07 & 15.85 & 14.63 & 15.54 & \underline{17.29} & \textbf{19.04} & 10.1\% \\
\cmidrule(lr){2-16}
    \multirow{5}{*}{\rotatebox{90}{\textbf{Kitchen}}} & MRR & 3.79 & 4.24 & 3.60 & 3.99 & 4.10 & 4.65 & 4.78 & 4.89 & 4.78 & 4.70 & 4.98 & \underline{5.04} & \textbf{5.56} & 10.3\% \\
     & N@5 & 3.35 & 3.22 & 3.90 & 4.25 & 3.68 & 4.16 & 4.37 & 4.28 & 4.12 & 5.10 & 4.52 & \underline{5.44} & \textbf{6.02} & 10.7\% \\
     & N@10 & 3.93 & 5.66 & 4.55 & 4.86 & 4.32 & 4.94 & 5.08 & 5.16 & 5.28 & 5.75 & 5.31 & \underline{6.37} & \textbf{7.01} & 10.0\% \\
     & H@5 & 4.78 & 4.79 & 5.62 & 5.81 & 5.17 & 5.74 & 6.06 & 6.22 & 6.07 & 7.12 & 6.22 & \underline{7.85} & \textbf{8.65} & 10.1\% \\
     & H@10 & 6.62 & 5.66 & 7.67 & 7.70 & 7.15 & 8.18 & 8.27 & 8.37 & 8.58 & 9.12 & 8.67 & \underline{10.73} & \textbf{11.74} & 9.4\% \\
    \bottomrule
  \end{tabular}
  \centering
\caption{Performance comparison across two cross-domain scenarios. The best results are highlighted in \textbf{bold}, and the second-best are \underline{underlined}. \textbf{Imp.} indicates the relative improvement percentage of our model over the strongest baseline, with statistical significance confirmed via a t-test at $p < 0.05$ compared to the best-performing baseline.}

  \label{tab:transposed_results}
\end{table*}

%% file: Content/05-Experiment.tex
\section{Experiment}
 This section addresses the following research questions:
\textbf{RQ1: }How does DPG-Diff perform compared to existing CDSR models? \textbf{RQ2: }What is the contribution of each component in DPG-Diff, particularly the preference-guided diffusion module, to its overall performance? \textbf{RQ3: }Is DPG-Diff robust to noisy interactions, and how does it maintain stability under such conditions? \textbf{RQ4: }What is the computational cost of DPG-Diff compared to other CDSR models?

\normalsize

We begin by detailing the experimental setup, including the evaluation protocol, datasets, and implementation details. Subsequently, we present and analyze the results to assess the performance of DPG-Diff against baselines.

\subsection{Experiment settings}

\subsubsection{Datasets}
We follow the experimental settings of C2DREIF~\cite{C2DREIF}, including datasets, protocols, and baselines, using two CDSR benchmarks (Movie-Book and Food-Kitchen) from Amazon. Users and items with fewer than 10 interactions are removed. We further exclude users who interacted with only one domain or fewer than three items per domain. A leave-one-out strategy is applied: the last interaction is used for testing, the second-to-last for validation, and the rest for training. Table~\ref{tab:dataset_stats} summarizes the filtered dataset statistics.

\input{Fig/stat}

To ensure an unbiased evaluation, we follow the standard protocol of sampling 1,000 items per test case, including 1 positive and 999 negative items, as commonly adopted in prior work~\cite{rendle}.
Top-K performance is reported using MRR (Mean Reciprocal Rank at 10), N@{5,10} (Normalized Discounted Cumulative Gain), and H@{5,10} (Hit Ratio).

\subsubsection{Baseline Comparison}
We compare DPG-Diff against the following baselines, including:

\textit{\textbf{Sequential Recommendation (SR):}}
SASRec~\cite{SASrec}: A self-attentive model capturing long-range dependencies. TCSR(TiCoSeRec)~\cite{TiCoSeRec}: Enhances CoSeRec with temporal cycle modeling. Diffu(DiffuRec)~\cite{diff4serec}: A diffusion-based sequential recommendation model.

\textit{\textbf{Cross-Domain Sequential Recommendation (CDSR):}}
CGRec: A contrastive graph-based sequential recommendation model. PSJ(PSJNet)~\cite{PSJNet}: Transfers user intentions via a parallel split-join scheme. C2DSR~\cite{C2DSR}: Leverages GNNs and InfoMax loss for cross-domain alignment. PCDSR~\cite{PCDSR}: Incorporates proxy items and temporal intervals. EAGCL~\cite{EAGCL}: Uses external attention and memory-sharing MLPs. LEAGCN~\cite{LEAGCN}: Captures collaborative filtering signals across domains. ABXI~\cite{ABXI}: Introduces invariant interest adaptation using LoRA-based modules. C2DREIF~\cite{C2DREIF}: Combines contrastive learning with disentangled representations. DREAM ~\cite{DREAM}: A recent CDSR model leveraging decoupled extraction and contrastive learning mechanisms.

\subsubsection{Implementation Details}
Our model is implemented in PyTorch~\cite{paszke2019pytorch}, following the same evaluation protocol and hyperparameter settings as C2DREIF~\cite{C2DREIF} for fair comparison. The embedding size and mini-batch size are set to 512. We train for 100 epochs using the Adam optimizer~\cite{adam}.

\subsection{Performance Comparison}
\textbf{Overall Performance Summary (RQ1)}
DPG-Diff consistently delivers superior performance across all evaluated cross-domain datasets and metrics, establishing itself as a robust and generalizable solution for CDSR. We summarize the key observations from our experiments as follows: \textbf{Cross-domain sequential recommendation models} consistently outperform traditional single-domain baselines, demonstrating the benefit of leveraging cross-domain signals to enhance knowledge transfer and user modeling. \textbf{DiffuRec}, a diffusion-based sequential recommender, not only surpasses standard sequential recommendation models but also outperforms several early CDSR approaches. This highlights the potential of diffusion modeling to capture complex and entangled user behavior across domains. \textbf{C$^2$DREIF}, \textbf{C$^2$DSR}, and \textbf{DREAM} adopt a decoupled learning strategy that separates cross- and single-domain modeling. By incorporating contrastive learning techniques to align sequences across domains, these models achieve promising results, showcasing the value of disentangled representation learning and contrastive objectives in CDSR. \textbf{DPG-Diff} further advances performance by integrating preference-guided diffusion and domain-specific adaptation. DPG-Diff achieves the highest Hit@10 score on the Movie dataset, outperforming the strongest baseline (DREAM) by 28.2\%, demonstrating its effectiveness in capturing transferable and personalized user preferences.


\input{Fig/Ablation}

\textbf{Ablation Study (RQ2)}
Table~\ref{tab:transposed_results} presents an ablation study evaluating the contribution of each component in DPG-Diff. Starting from the base diffusion model (Diff), we observe incremental improvements with each added module:
\begin{itemize}
    \item \textbf{Diff + DE}: Adding the Disentangled Encoder improves NDCG@10 from 5.95 to 5.83 on the Movie dataset, indicating that domain-aware encoding helps capture more relevant user signals.
    \item \textbf{Diff + DE + G}: Incorporating Preference Guidance further boosts performance to 8.34, demonstrating its critical role in leveraging preferences in the diffusion process.
    \item \textbf{Diff + DE + Tri-CL}: Combining DE and Tri-CL yields stronger results, but still falls short of the full model.
    \item \textbf{DPG-Diff (Full)}: The complete model achieves the highest scores across all domains, confirming the synergistic effect of diffusion modeling, disentangled preference guidance, and contrastive alignment.
\end{itemize}

This step-by-step comparison highlights the complementary nature of each component and validates their individual and collective contributions to CDSR.

\textbf{Robust Study (RQ3)}
To evaluate the robustness of DPG-Diff against noisy interactions, such as contextually irrelevant items and conflicting preferences. We conduct a controlled noise injection experiment. Specifically, we simulate noise by randomly inserting and substituting items within test sequences, gradually increasing the noise rate during inference. Both DPG-Diff and DREAM are trained on the same dataset and evaluated under identical noisy conditions.

\begin{figure}
    \includegraphics[width=\linewidth]{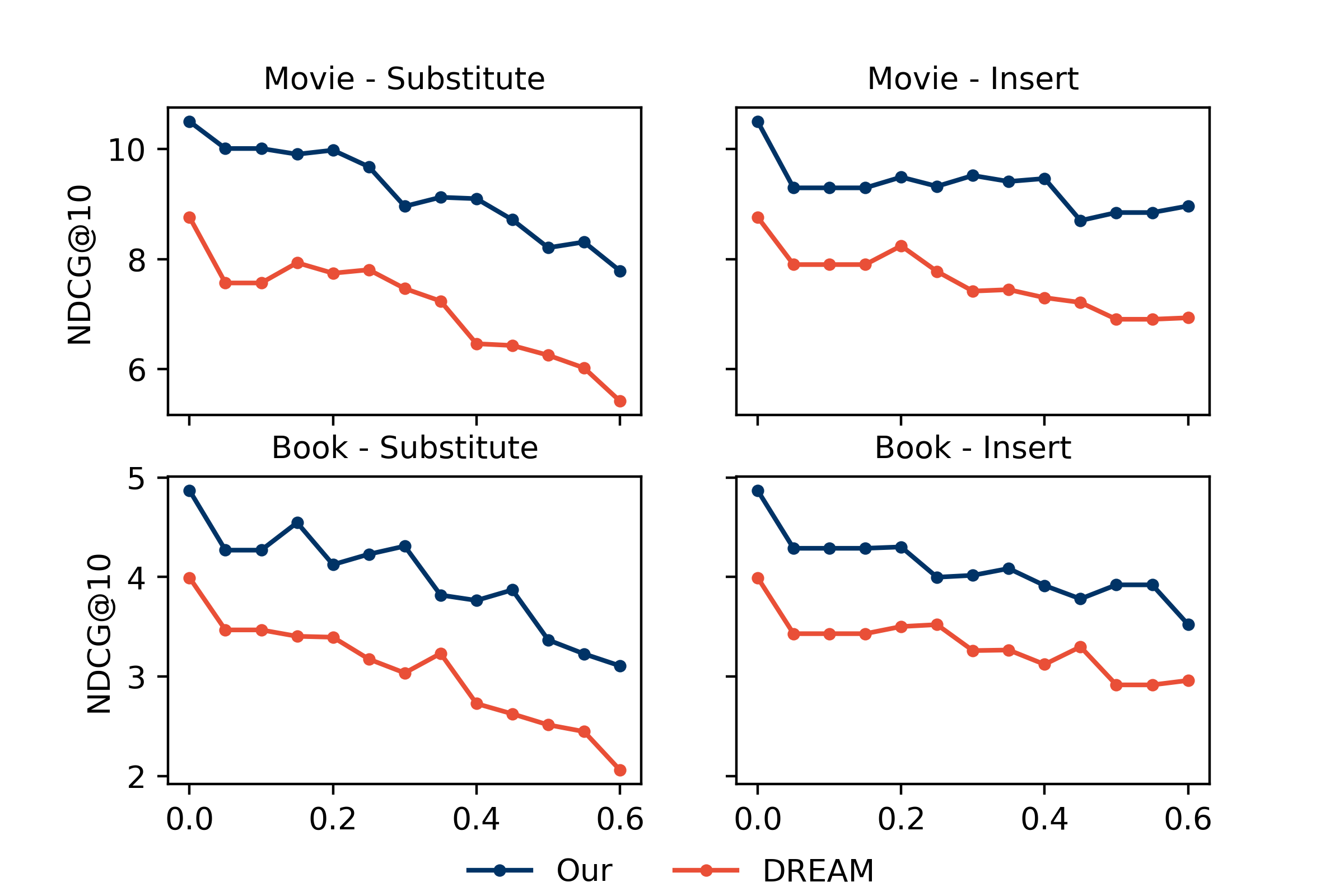}
\caption{Robustness comparison between DPG-Diff and DREAM under increasing noise rates.}
    \label{fig:noise}
\end{figure}

As shown in Figure~\ref{fig:noise}, DPG-Diff consistently outperforms DREAM across all noise levels and exhibits a slower degradation in performance. Notably, at a 30\% noise rate, DPG-Diff retains 85\% of its original performance NDCG@10 score, while DREAM falls below 70\%, underscoring the effectiveness of disentangled preference guidance in filtering out irrelevant signals and preserving recommendation quality under noisy conditions.

\textbf{Model Efficiency Analysis (RQ4)} Compared to DREAM, DPG-Diff achieves efficiency gains by simplifying its learning objectives. It replaces focal loss, which introduces additional overhead from dynamic probability weighting, with standard cross-entropy, reducing memory and time demand. Furthermore, DPG-Diff adopts an unsupervised tri-view CL instead of supervised contrastive learning, which typically requires more positive pairs and label-based pair selection, incurring $\mathcal{O}(b^2 \cdot d)$ complexity. By avoiding these costly components, DPG-Diff streamlines training as well as reduces memory and time usage while maintaining strong representation quality.

\begin{figure}
    \centering
    \includegraphics[width=\linewidth]{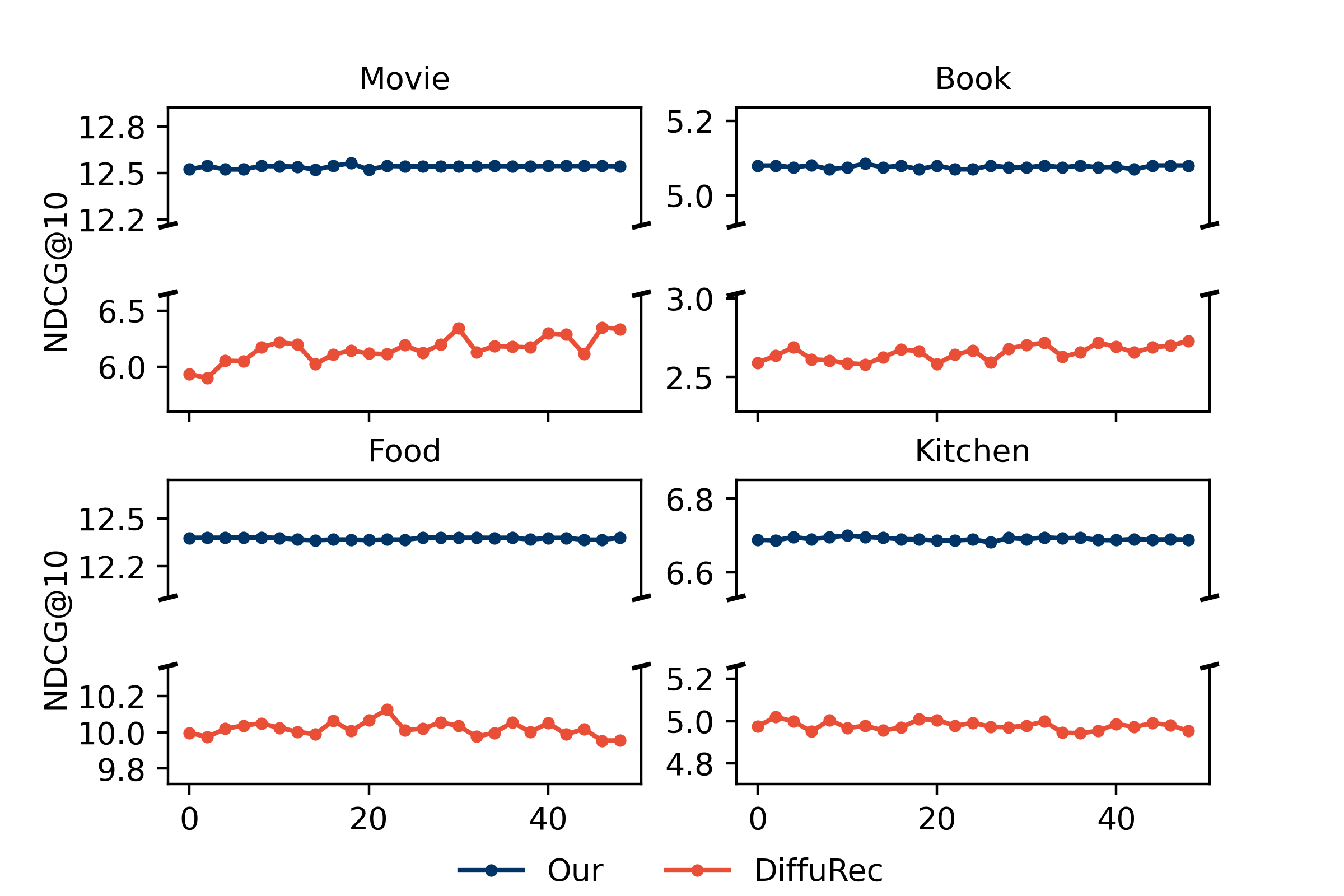}
    \caption{Effect of inference step of DPG-Diff and DiffuRec.}
    \label{fig:inference_step}
\end{figure}

As illustrated in Figure~\ref{fig:inference_step}, DPG-Diff consistently outperforms DiffuRec across all inference steps in terms of NDCG, achieving strong performance even in early stages at all datasets. This demonstrates its stability and efficiency, in contrast to DiffuRec’s slower and more fluctuating convergence. Therefore, our DPG-Diff shows superior model performance than DiffuRec while maintaining lower inference cost under different step settings.  The robustness of DPG-Diff stems from its disentangled preference guidance, enabling high-quality recommendations with fewer inference steps.

%% file: Fig/stat.tex
\begin{table}
\centering
\small
\begin{tabular}{lrr}
\toprule
\textbf{Statistic} & \textbf{Food / Kitchen} & \textbf{Movie / Book} \\
\midrule
\#Items (X / Y)         & 29,207 / 34,886 & 36,845 / 63,937 \\
Avg. Sequence Length    & 9.91            & 11.98            \\
\#Training Sequences    & 34,117          & 58,515           \\
\#Validation (X / Y)    & 2,722 / 5,451   & 2,032 / 5,612    \\
\#Test (X / Y)          & 2,747 / 5,659   & 1,978 / 5,730    \\
\bottomrule
\end{tabular}
\caption{Statistics of the cross-domain datasets.}
\label{tab:dataset_stats}
\end{table}

%% file: Fig/Ablation.tex
\begin{table}
\centering

\small
\begin{tabular}{lrrrr}
\hline
\textbf{Variants} & \textbf{Food} & \textbf{Kitchen} & \textbf{Movie} & \textbf{Book}\\
\hline
Diff & 9.06 & 4.21 & 5.95 & 2.36 \\
Diff + DE & 9.24 & 4.27 & 5.83 & 3.67 \\
Diff + DE + G & 10.43 & 5.38 & 8.34 & 3.83 \\
Diff + DE + Tri-CL & 10.44 & 4.90 & 9.07 & 4.01 \\
\textbf{DPG-Diff (Full)} & \textbf{13.50} & \textbf{5.58} & \textbf{13.50} & \textbf{5.58} \\
\hline
\end{tabular}
\caption{Model variants for Ablation Study.}
\label{tab:transposed_results}
\begin{minipage}{0.95\linewidth}
\small
\textbf{Abbreviations:} \textbf{DE} = Disentangled Encoder; \textbf{G} = Disentangled Preference Guidance; \textbf{Tri-CL} = Tri-view Contrastive Learning.  
\textbf{Diff} refers to the base diffusion-based sequential recommender.
\end{minipage}
\end{table}

%% file: Content/06-Relatedwork.tex
\section{Related Work}
Recent recommendation systems have explored both cross-domain learning and generative modeling. We review related work in two key areas: Cross-Domain Sequential Recommendation and Diffusion Models in Recommendation.

\textbf{Cross-Domain Sequential Recommendation}  (CDSR) aims to improve recommendation performance by leveraging user sequential behaviors across multiple domains. Models like PSJNet~\cite{PSJNet} and $\pi$-Net~\cite{pinet} use gating mechanisms for knowledge transfer. DA-GCN~\cite{DAGCN}, C2DSR~\cite{C2DSR}, and C2DREIF~\cite{C2DREIF} extend GNNs to encode sequential dependencies. Transformer-based approaches such as RecGURU~\cite{RecGURU} and DREAM~\cite{DREAM} model multi-domain sequences, while ABXI~\cite{ABXI} introduces low-rank adaptation for domain-invariant interest learning. Despite these advances, many CDSR models struggle to disentangle domain-specific signals and mitigate cross-domain noise, limiting their generalization capabilities. 

\textbf{Diffusion Models in Recommendation} Diffusion models (DMs), inspired by non-equilibrium thermodynamics, generate data by reversing a noise process~\cite{YangZSHXZZCY24}, guided by conditional inputs~\cite{DhariwalN21, RombachBLEO22}. In recommendation tasks, user behavior sequences serve as conditions to generate item embeddings. DiffuRec~\cite{diff4serec}, DCRec \cite{huang2024dual}, and DreamRec~\cite{DreamRec} apply DMs to sequential recommendation, but often compress complex behaviors into a single guidance vector, which limits expressiveness. To address cross-domain challenges, DMCDR~\cite{DMCDR} and CDCDR~\cite{CDCDR} adapt DMs for cross-domain recommendation by generating target-domain representations or learning unified item distributions. However, these models treat user histories as unordered sets, neglecting temporal dynamics, making them unsuitable for CDSR tasks.



%% file: Content/07-Conclusion.tex
\section{CONCLUSION}
In this paper, we proposed \textbf{DPG-Diff}, the first diffusion-based framework tailored for Cross-Domain Sequential Recommendation. DPG-Diff leverages a disentangled preference-guided diffusion process to generate personalized user representations by conditioning on interaction histories from both source and target domains. Extensive experiments demonstrate that DPG-Diff achieves state-of-the-art performance across multiple benchmarks. As future work, we plan to extend our framework to multi-domain scenarios to explore its scalability and generalization.

%% file: Fig/Algo_Inference.tex
\begin{algorithm}[ht]
\small
\caption{Inference Procedure of DPG-Diff}
\label{alg:inference}
\begin{algorithmic}[1]
\REQUIRE User interaction sequences \( s_x, s_y \)
\REQUIRE Item embeddings \( E_x, E_y \); diffusion steps \( T \)
\ENSURE Predicted scores \( \hat{y}_x, \hat{y}_y \)

\STATE \textbf{Encode single-domain preferences:}
\[
\hat{g}_x = \text{Encoder}_x(s_x), \quad \hat{g}_y = \text{Encoder}_y(s_y)
\]

\STATE \textbf{Fuse representations:} Combine \( \hat{g}_x \) and \( \hat{g}_y \) into disentangled guidance \( \hat{g}_d \)

\STATE \textbf{Initialize diffusion:} Sample noise \( x_T \sim \mathcal{N}(0, I) \)

\FOR{\( t = T \) \textbf{to} \( 1 \)}
    \STATE \textbf{Denoise step:} Estimate \( x_{t-1} \) via
    \[
    p_{\theta}(x_{t-1} \mid x_t) = \mathcal{N}(x_{t-1}; \mu_{\theta}(x_t, t), \Sigma_{\theta}(x_t, t))
    \]
\ENDFOR

\STATE \textbf{Final representation:} \( \hat{x}_0 = x_0 \)

\STATE \textbf{Compute prediction scores:}
\[
\hat{y}_x = \text{softmax}(\hat{x}_0^\top E_x + \hat{g}_x^\top E_x)\]
\[
\hat{y}_y = \text{softmax}(\hat{x}_0^\top E_y + \hat{g}_y^\top E_y)
\]

\STATE \textbf{Return:} Top-\( k \) items from \( \hat{y}_x \) and \( \hat{y}_y \)
\end{algorithmic}
\normalsize
\end{algorithm}